\documentstyle[aps,preprint]{revtex}

\newcommand{\chpb}{\mbox{$\chi_\perp$}}
\newcommand{\chp}{\mbox{$\tilde{\chi}_\perp$}}
\newcommand{\phidot}{\mbox{$\dot{\phi}$}}
\newcommand{\thdot}{\mbox{$\dot{\theta}$}}

\newcommand{\kpa}{\mbox{$K_{||}$}}
\newcommand{\phidott}{\mbox{$\dot{\phi}^2$}}

\newcommand{\phiddot}{\mbox{$\ddot{\phi}$}}
\newcommand{\thddot}{\mbox{$\ddot{\theta}$}}
\newcommand{\dphi}{\mbox{$\delta \phi$}}
\newcommand{\dth}{\mbox{$\delta \theta$}}
\newcommand{\phibar}{\mbox{$\bar{\phi}$}}
\newcommand{\thbar}{\mbox{$\bar{\theta}$}}
\newcommand{\ben}{\begin{equation}}
\newcommand{\een}{\end{equation}}
\newcommand{\bena}{\begin{eqnarray}}
\newcommand{\eena}{\end{eqnarray}}
\newcommand{\nutil}{\mbox{$\tilde{n}_u$}}
\newcommand{\nutilt}{\mbox{$\tilde{n}_{u}^{2}$}}

\begin{document}
\draft

\title{
First- and second-order transitions
of the escape rate in ferrimagnetic or antiferromagnetic particles
}

\author{
Gwang-Hee Kim
}

\address{Department of Physics, Sejong University,
Seoul 143-747, Republic of Korea}
\date{Received \hspace*{10pc}}
\maketitle

\thispagestyle{empty}

\begin{abstract}
Quantum-classical escape-rate transition has been studied
for two general forms of magnetic anisotropy in ferrimagnetic
or antiferromagnetic particles. 
It is found that the range of the first-order transition is greatly reduced
as the system becomes ferrimagnetic and  there is no 
first-order transition in almost
compensated antiferromagnetic particles.
These features can be tested experimentally
in nanomagnets like molecular magnets.
\\
\end{abstract}

Understanding the mechanism of transition between two states in a bistable
system is of utmost importance of the fast-growing area of macroscopic
quantum phenomena (MQP).\cite{leg}
The rf SQUID and the closely related current-biased Josephson
junction\cite{leg87} have been typical examples of MQP. 
In recent years, as another good candidates for MQP,
small magnetic particles have been investigated via
quantum tunneling of magnetization.\cite{gun}
In general, transition between two states can occur
due to two mechanisms. At sufficiently high temperature the transition rate 
$\Gamma$ obeys the Arrhenius laws, $\Gamma \sim \exp (-U/k_BT)$,
with $U$ being the height of the energy barrier.
At a temperature low enough to ignore the thermal fluctuation
quantum tunneling is relevant with $\Gamma \sim \exp (-U/\hbar \omega)$
where $\omega$ is the oscillation frequency in the well. Thus, the 
crossover temperature $T_0$ is expected to exist between
the thermal activation and quantum tunneling. Whether
the two regimes smoothly join or not has been intensive subject over the past
few years.\cite{chu92,gor,chu,lmp,kim,mul}

The criterion of  first- or  second-order transitions for the escape
rate has been suggested by Chudnovsky,\cite{chu92} who showed that
if the oscillation period $\tau(E)$ of the instanton is monotonic (nonmonotonic)
function where $E$ is the energy of the instanton,
$d\Gamma(T)/dT$ becomes continuous (discontinuous), i.e., the
second- (the first-) order transition around the
crossover temperature. Since then, Gorokhov and 
Blatter\cite{gor} carried out
the nonlinear perturbation near the top of the barrier, obtained
a criterion for the first-order transition in tunneling problem based on
Ref. \cite{chu92}, and applied it to the phase transition of vortex
creep pinned in the columnar defect. Later, theoretical
investigations for the transition in spin systems have been
performed by several groups.\cite{chu,lmp,kim,mul} 
Until now theoretical studies have been
focused on the ferromagnetic particle. However, considering that most
ferromagnetic systems are actually ferrimagnetic particles,
the strong exchange interaction should be taken
into consideration. 
The exchange interactions are expected
to suppress the first-order transition, and thereby, if it is large,
there may be no first-order transition. Thus, it is clearly
desirable to assess its importance for the phase transition before
contemplating real experiment.
Indeed, the exchange interaction in the Hamiltonian does not permit
a simple mapping onto the particle problem\cite{chu,kim}  and
the periodic instantons in  the reduced one-dimension.\cite{lmp,mul}
None of such previous theoretical methods are
applicable to the present spin system with the exchange interaction.
In this paper we will develop new approach to treat the
phase transition in ferrimagnetic or antiferromagnetic particles
based on  the spin-coherent-state path integral,
and present complete analytic results of the phase boundary between
first- and second-order transitions for two general forms of the
magnetic anisotropy energy. Also, we will discuss effects of the 
fourth order anisotropy in the Hamiltonian on the phase transition 
in molecular clusters.\cite{cac}

Let us consider a small ferrimagnetic or antiferromagetic particle 
with two magnetic sublattices whose magnetizations,
${\bf m_1}$ and ${\bf m_2}$,
are coupled by the strong exchange interaction 
${\bf m_1}\cdot {\bf m_2}/\chpb$, where $\chpb$ is
the perpendicular succeptibility.
In the case of a non-compensation of sublattices with
$m(=m_1 - m_2 >0) $, the Euclidean action can be taken to be\cite{chu95}
\begin{eqnarray}
S_E (\theta,\phi) &=& V \int d\tau
\left( i{ m_1+m_2 \over \gamma} \phidot-i{m \over \gamma} 
\phidot \cos \theta + {\cal L}_{1 E} \right),  \label{action}
\end{eqnarray}
where $V$ is a volume of the particle, 
$\gamma$ the gyromagnetic ratio, and
$\theta$, $\phi$ spherical coordinates of sublattice magnetization
${\bf m_1}$ which determines the direction of the N\'eel vector.
Here, ${\cal L}_{1 E}$ is the Euclidean Lagrangian density which
is composed of the exchange interaction, the anisotropy energy and 
the energy given by an external magnetic field.
The first term in the action (\ref{action}) known as 
the topological term\cite{los}
generates the phase factor related with
quantum phase interference. In this work we do
not include this term because it is not the purpose of this paper
to discuss this issue.

In the simplest easy-axis anisotropy
along the $z$-axis and a transverse magnetic field along the $x$-axis 
the Euclidean Lagrangian density is expressed as\cite{iva}
\bena
{\cal L}_{1E}&=&{\chp \over 2 \gamma^2}[(\dot{\theta}+i \gamma H \sin \phi)^2
+ (\dot{\phi}\sin\theta+i\gamma H\cos \theta \cos \phi)^2 ]
\nonumber \\
&&\ \ \ +\kpa \sin^2 \theta-mH\sin \theta \cos \phi,
\label{landen}
\eena
where $\chp=\chpb (m_2/m_1)$.
Then, the corresponding classical trajectory  is determined by the equations
\begin{eqnarray}
&-&i n_u  \phidot \sin \theta + x_u
[ \thddot - {1 \over 2} \phidott \sin 2\theta +2 i b
\phidot \cos \phi \sin^2 \theta 
\nonumber \\
&&-{1 \over  2}b^2  \sin 2\theta \cos^2\phi ]
 -2 \cos \theta(\sin \theta-h \cos \phi)=0, \label{phiequ}\\
&&i n_u \thdot +x_u
[ \phiddot \sin \theta + 2 \phidot \thdot \cos \theta  
-2 i b \thdot \cos \phi \sin \theta 
\nonumber \\
&&\ \ \ +{1 \over  2}b^2  \sin 2\phi \sin  \theta ]
-2h \sin \phi=0. \label{thequ}
\end{eqnarray}
where we have  introduced $m/(\kpa \gamma)=n_u$, 
$\chp/(\kpa \gamma^2)=x_u$, $\gamma H=b$,
$H/H_c=h$, and $H_c =2 \kpa/m$. Note that
$H_c$ is the critical magnetic field at which the barrier vanishes.
In this case we decompose
$\theta$ ($\phi$) into the position of the top of the barrier 
$\thbar$ ($\phibar$) and a fluctuation term $\dth$ ($\dphi$), i.e.,
$\theta=\thbar+\dth$ $(\phi=\phibar+\dphi)$
for the behavior of the weakly time-dependent solutions,
where $\thbar=\pi/2$ and $\phibar=0$. 
Our goal is to solve Eqs. (\ref{phiequ}) and (\ref{thequ}) 
for $\dth(\tau)$ and $\dphi(\tau)$
and find the correction to the oscillation period away from the
thermal saddle point.\cite{gor}
Denoting 
$\delta {\bf{\Omega}}(\tau) \equiv (\dth(\tau)$, $\dphi(\tau))$, we have
$\delta {\bf{\Omega}}(\tau+\beta \hbar)=\delta {\bf{\Omega}}(\tau)$
at finite temperature and write it as Fourier series 
$\delta {\bf{\Omega}}(\tau)=
\sum^{\infty}_{n=-\infty}\delta {\bf{\Omega}}_n \exp(i\omega_n \tau)$
where $\omega_n=2\pi n/\beta \hbar$. Since simple analysis shows that
$\dth$ is real and $\dphi$ imaginary in this model, to lowerst order we
write them in the form $\dth \simeq a  \theta_{u1} \cos (\omega \tau)$ and 
$\dphi \simeq i a \phi_{u1} \sin (\omega \tau)$ in the vicinity of the thermal
saddle-point solution. Here we have parametrized the solutions of the equation
of motion by the amplitude $a$ of the oscillations, which quantifies the
difference between the thermal and the time-dependent solutions near
the top of the barrier.\cite{gor} 
Substituting them into Eqs. (\ref{phiequ}) and (\ref{thequ}), we
obtain the relation 
\ben
{\phi_{u1} \over \theta_{u1}} ={ x_u (\omega^{2}_{u \pm} -b^2) -2
+2h \over \nutil  \omega_{u \pm}}
={-\nutil \omega_{u \pm} \over x_u (\omega^{2}_{u \pm}-b^2) +2 h},
\een
where $\tilde{n}_u=n_u-2bx_u$, and the oscillation frequency
\bena
&&\omega^{2}_{u\pm} =b^2 - {\nutilt -2(1-2h) x_u  \over 2   x^{2}_{u}} \nonumber \\
&&\pm { \sqrt{[\nutilt -2(1-2h) x_u]^2+16h(1-h)x^{2}_{u}-4 b\nutilt x^{2}_{u} } 
\over 2   x^{2}_{u} }.
\eena

In order to see the behavior of the oscillation period via the frequency
we need to investigate Eqs. (\ref{phiequ}) and (\ref{thequ})
by writing $\dth \simeq a  \theta_{u1} \cos (\omega \tau) 
+ \delta \theta_2$ and 
$\dphi \simeq i a \phi_{u1} \sin (\omega \tau)+ i \delta \phi_2$, where
$\delta \theta_2$ and $\delta \phi_2$ are $O(a^2)$.
Neglecting terms of order higher than $a^2$, we find 
$\omega=\omega_{u+}$, $\delta \theta_2=0$, and $\delta \phi_2 =0$. This implies
that there is no shift in the oscillation frequency. Thus, to third order in perturbation
theory we use $\dth \simeq a  \theta_{u1} \cos (\omega \tau) + \delta \theta_3$ and 
$\dphi \simeq i a \phi_{u1} \sin (\omega \tau)+ i \delta \phi_3$, where
$\delta \theta_3$ and $\delta \phi_3$ are $O(a^3)$. Substituting them into
Eqs. (\ref{phiequ}) and (\ref{thequ}), and neglecting terms of order higher than $a^3$, 
we obtain for the change of the oscillation frequency
\bena
n^{4}_{u} y^{2}_{u} (\omega^2 -\omega^{2}_{u+})(\omega^2 -\omega^{2}_{u-})=
a^2{\theta^{2}_{u1} \over 4} g_u(h,y_u).
\label{omequ}
\eena
Here $y_u=x_u /n^{2}_{u} (=\chp\kpa/m^2)$ 
indicates the relative magnitude of the noncompensation.
For large noncompensation ($y_u \ll 1$, i.e., 
$m \gg \protect\sqrt{\tilde{\chi}_{\perp} K_{||}}$) and for small
noncompensation ($y_u \gg 1$, i.e., 
$m \ll \protect\sqrt{\tilde{\chi}_{\perp} K_{||}}$), the system
becomes ferromagnetic and nearly compensated
antiferromagnetic, respectively.
Without any approximation, $g_u(h,y)$ in Eq. (\ref{omequ}) is so
complicated that it is not illuminating to present its specific form. 
Noting that  the first-order region is  expected to shrink by increasing $y_u$, it
is slightly influenced by the magnetic field terms in the exchange interaction
of Eq. (\ref{landen}). Thus, neglecting those terms in the exchange interaction, 
we can obtain the analytic form of $g_u(h,y)$ approximaterly given by
\begin{eqnarray}
g_u(h,y)&\simeq&{1 \over y^2}[2-2y+12 hy-3 y^2 +8 hy^2 -2 y^3 \nonumber \\
&-&(2+2y+4hy-y^2) \sqrt{1-4y+8hy+4 y^2}].
\label{appxgu}
\end{eqnarray}
As shown by Chudnovsky,\cite{chu92} 
if the oscillation period $\tau$ is not a monotonic
function of $a$ where $a$ is a function of $E$ in the absence of dissipation, 
the system exhibits a first-order transition. Thus, the period $\tau(=2\pi/\omega)$
in Eq. (\ref{omequ}) should be less than $\tau_{u+} (=2 \pi/\omega_{u+})$, i.e.,
$\omega > \omega_{u+}$ for the first-order transiton. It implies that
$g_u(h,y_u) >0$  in Eq. (\ref{omequ}) for the first-order
transition, and  $g_u(h,y_u)=0$ determines the phase
boundary between the first- and the second-order transition. 
Using Eq. (\ref{appxgu}), we approximaterly get the phase boundary
\bena
h(y_u) \simeq {2+y_u \over 6}-{12+20y_u+44 y^{2}_{u}-y^{3}_{u} \over 12 f(y_u)^{1/3}} +
{f(y_u)^{1/3} \over 12 y_u},
\label{appxphau}
\eena
where
\bena
&f&(y)=-18y^2 -350 y^3 -372 y^4 +66 y^5 -y^6 \nonumber \\
&&+6  \sqrt{3y^3 (1+3y+y^2)^2 (16-13y+328y^2-6y^3)}.
\eena
Also, in  order to compare Eq.  (\ref{appxphau}) with the phase boundary
from the exact calculation, we have employed the numerical method for
the exact phase boundary, whose results are illustrated in Fig. \ref{phaseu}.  
The approximate result (\ref{appxphau}) is found to be
slightly different from the exact one obtained by the numerical methods.
The scaled magnetic field, $h$ for the phase boundary
decreases linearly for $y_u \ll 1$, as $h \simeq 1/4 - y_u$ and
parabolically for $y_u \lesssim 1/2$, as $h \simeq 0.397 (y_u-1/2)^2$. Thus,
it is evident that the region for the first order transition is greatly reduced
as the system becomes ferrimagnetic and there is no first-order transition
in almost compensated antiferromagnetic particles.

We now turn to the biaxial symmetry with the Euclidean Lagrangian density
\bena
{\cal L}_{1E}&=&{\chp \over 2 \gamma^2}[\dot{\theta}^2
+ (\dot{\phi}\sin\theta)^2 ]
\nonumber \\
&&+K_1 \cos^2 \theta +K_2  \sin^2 \theta \sin^2 \phi,
\label{eneb}
\eena
where $K_1 > K_2 >0$ are the anisotropy constants. 
This model describes a hard axis $z$ and an easy
axis $x$, and the corresponding classical trajectory satisfies
\begin{eqnarray}
&-&i n_b  \phidot \sin \theta + x_b
\left( \thddot - {1 \over 2} \phidott \sin 2\theta \right) \nonumber \\
&&\ \ \ \ \ \ \ \ \ +\sin 2 \theta (1-k \sin^2 \phi)=0, \label{phieqb}\\
&&i n_b \thdot +x_b
\left( \phiddot \sin \theta + 2 \phidot \thdot \cos \theta  \right)\nonumber \\
&&\ \ \ \ \ \ \ \ \
-2k \sin \theta \sin \phi \cos \phi=0. \label{theqb}
\end{eqnarray}
where we have  introduced $m/(K_1\gamma)=n_b$, $\chp/(K_1 \gamma^2)=x_b$, 
and $K_2/K_1=k$. In the high temperature regime the solution of Eqs.
(\ref{phieqb}) and (\ref{theqb}) is $\thbar=\phibar=\pi/2$. 
In order to find the order of the transition, let us expand Eqs.
(\ref{phieqb}) and (\ref{theqb}) into a series around this solution,
as $\theta=\pi/2+\dth$ and $\phi=\pi/2+\dphi$. Simple analysis for
Eqs. (\ref{eneb})-(\ref{theqb}) shows that
$\dth$ is imaginary and $\dphi$ real. Thus, to lowest order in
perturbation theory, we write $\dth \simeq i a \theta_{b1} \sin (\omega \tau)$
and $\dphi \simeq a \phi_{b1} \cos (\omega \tau)$. Substituting them
into Eqs. (\ref{phieqb}) and (\ref{theqb}) while neglecting terms of order
higher than $a$, we have
\ben
{\phi_{b1} \over \theta_{b1}} ={ x_b \omega^{2}_{b \pm} 
+2 (1-k) \over n_b \omega_{b \pm}}
=-{n_b \omega_{b \pm} \over x_b \omega^{2}_{b \pm} -2 k}
\een
and the oscillation frequency
\bena
\omega^{2}_{b\pm} &=&- {n^{2}_{b}  +2(1-2k)x_b \over 2   x^{2}_{b}} \nonumber \\
&&\pm { \sqrt{n^{4}_{b} +4(1-2k) n^{2}_{b} x_b + 4 x^{2}_{b}} 
\over 2   x^{2}_{b} }.
\eena

Next, let us write $\dth \simeq i a \theta_{b1} \sin (\omega \tau)+i \delta
\theta_2$, and $\dphi \simeq a \phi_{b1} \cos (\omega \tau) +\delta \phi_2$,
where $\delta \theta_2$ and $\delta \phi_2$ are of the order of $a^2$.
Inserting them  into  Eqs. (\ref{phieqb}) and (\ref{theqb}),
we arrive at $\omega =\omega_{b+}$ and $\delta \theta_2=
\delta \phi_2=0$. In order to find the change of the oscillation
period, we proceed to the third order of perturbation theory by
writing $\dth \simeq i a \theta_{b1} \sin (\omega \tau)+i \delta
\theta_3$, and $\dphi \simeq a \phi_{b1} \cos (\omega \tau) +\delta \phi_3$.
Substituting them again into Eqs. (\ref{phieqb}) and (\ref{theqb}) and retaining
only terms up to $O(a^3)$, we have
\bena
n^{4}_{b} y^{2}_{b} (\omega^2 -\omega^{2}_{b+})(\omega^2 -\omega^{2}_{b-})=
a^2{\phi^{2}_{b1} \over 4} g_b(k,y_b),
\label{omeqb}
\eena
where $y_b=x_b /n^{2}_{b} (=\chp K_1/m^2)$ and
\begin{eqnarray}
g_b(k,y)&=&{2 \over y^2}[1+3y-4ky-4 y^3 \nonumber \\
&-&(1+y-2 y^2) \sqrt{1+4y-8ky+4 y^2}].
\end{eqnarray}
Here we note that the parameter $y_b$
has the same physical meaning as $y_u$ in
uniaxial symmetry by replacing $K_{||}$ by $K_1$.
Continuing in the present case as in the earlier model
we have $g_b(k,y_b) >0$ for the first-order transition and
the phase boundary given by
\ben
k={1\over 2} (1-y_b) (1+2 y_b)^2.
\een
As follows from Fig. {\ref{phaseb}, the criterion\cite{lmp} that 
in case of the biaxial symmetry 
the first-order transition  occurs at $k > 1/2$ 
is not valid in ferrimagnetic particles.
Fig. {\ref{phaseb} shows that the ratio of two anisotropy
constants, $k$ increases linearly for $y_b \ll 1$, as
$k \simeq (1+3y_b)/2$ and has a maximum at $y_b=1/2$.
Therefore, the region for the first-order transition is
greatly suppressed with increasing $y_b$ and vanishes
beyond $y_b=1/2$.

To illustrate the above results with concrete examples we
discuss two molecular clusters, ${\rm Mn_{12}}$ and ${\rm Fe_8}$.
Actually, both samples are ferrimagnetic, and thereby $y_u$
and $y_b$ should be taken into account in uniaxial and
biaxial symmetry, respectively. Taking the measured value
of the anisotropy parameter, e.g., $k=0.728$ for ${\rm Fe_8}$,
it is seen from Fig. \ref{phaseb} that the system may
exhibit the first-order transition for $y_b  \lesssim 0.157$ and the 
second-order transition beyond that region. This point should
be checked in experiment. 
Also, since neutron scattering data\cite{cac} shows
that an ${\rm Fe_8}$ (${\rm Mn_{12}}$) 
should be described by the biaxial (uniaxial) spin Hamiltonian
with the fourth order anisotropy term, it is meaningful to discuss how much it makes
an effect on the phase transition. 
Our analysis shows that the first-order region is slightly reduced by
adding the fourth order terms in the Hamiltonian. Taking  ${\rm Fe_8}$
for instance, the range of $y_b$ for the first-order transition is changed to
$0 \leq y_b \lesssim 0.153$, as shown in Fig. \ref{fth}. 
Thus, considering the fact that most ferromagnetic systems are ferrimagnetic,
it is very important to obtain the information on the
magnitude of the quantities $y_u$ and $y_b$ for observing the 
first-order transition in real experiments.

In conclusion, we have presented the phase diagrams
for first- and second-order transitions
in ferrimagnetic or antiferromagnetic particles.
It is found that the range of the first-order transition is greatly suppresed
by the exchange enhancement. In fact, this can be
qualitatively understood from the consideration that the exchange
interaction produces some effective magnetic field and thereby
plays an important role in reducing the range of
the first-order transition. In this respect, the first-order transition
can not be observed in almost compensated antiferromagnetic particles,
which is consistent with our analytic results. 
This general features
is expected to be observable in nanomagnets including molecular clusters.

Discussions with D. A. Gorokhov, W. Wernsdorfer  and B. Barbara
are acknowledged.
This work was supported 
by the Korea Research Foundation Grant (KRF-99-041-D00188)
\\

\begin{figure}
\caption{
Phase diagram $h(=H/H_c)$ vs $y_u(=\chp \kpa/m^2)$ obtained by the exact calculation
for the uniaxial symmetry in a transverse magnetic field. 
Note that $h$ and $y_u$ contain information of the magnetic field and the
relative noncompensation, respectively, and thereby
the system tends to be ferrimagnetic or antiferromagnetic
with increasing $y_u$. See the text for details. Inset: Comparison of the
exact  result (a) with the approximate one (b).
}
\label{phaseu}
\end{figure}

\begin{figure}
\caption{
Phase diagram $k(=K_2/K_1)$ vs $y_b(=\chp K_1/m^2)$ for the transition
in biaxial symmetry. Note that $k$ contains information 
of the relative strength of the biaxial two axes, and 
the first-order transition is expected for $k \protect\lesssim 1$ in which
the system starts to be biaxial from uniaxial.
Also, the parameter $y_b$ has the same physical meaning as $y_u$ in
uniaxial symmetry. See the text for details.
}
 \label{phaseb}
\end{figure}

\begin{figure}
\caption{
Change of the range of the first-order transition for the biaxial spin
Hamiltonian  in ${\rm Fe_8}$ (a) with and (b) without the fourth order 
anisotropy term in the Hamiltonian.
Note that (a) $0 < y_b \protect\lesssim 0.153$ and 
(b) $0< y_b \protect\lesssim 0.157$ for the
first-order transition.
}
 \label{fth}
\end{figure}

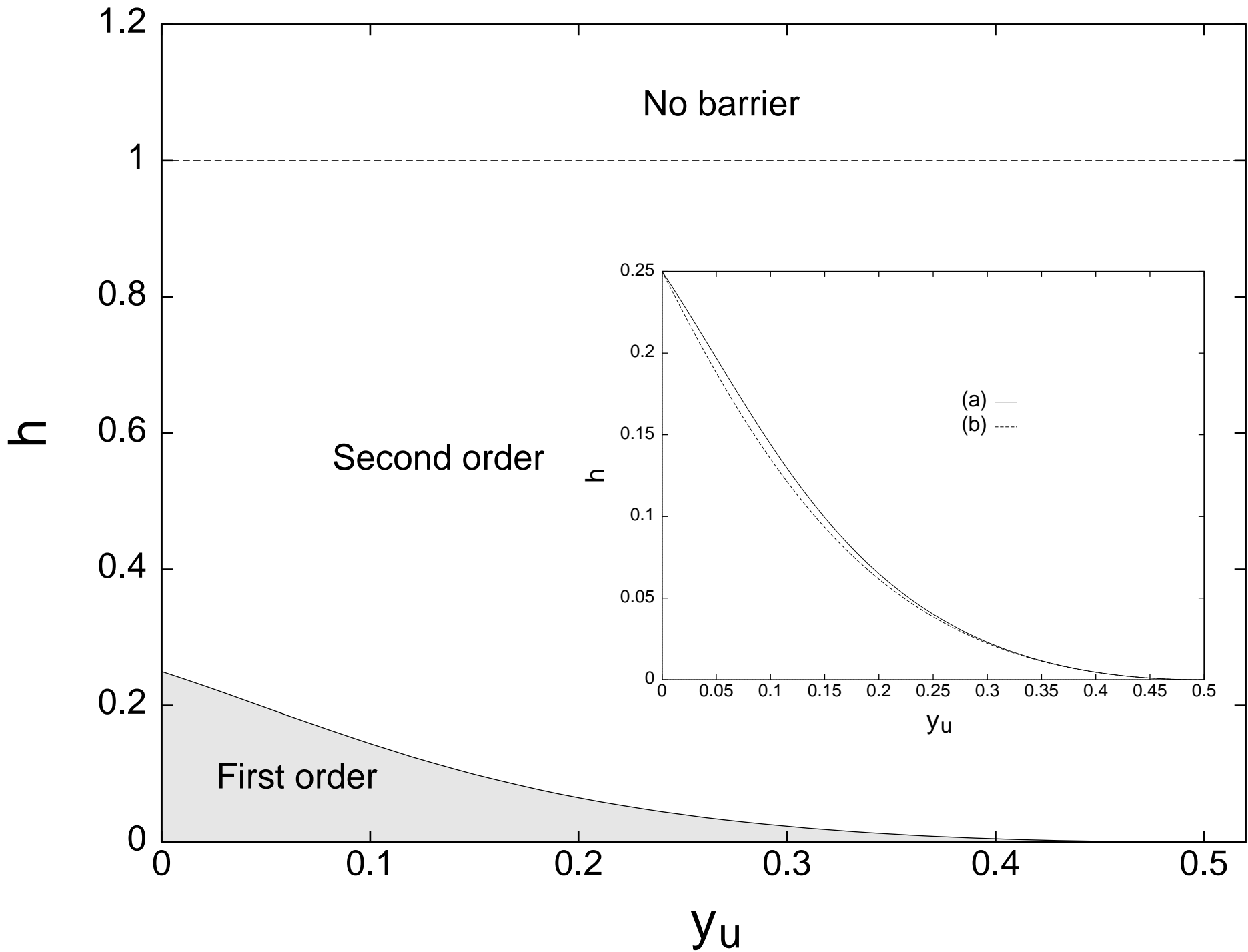
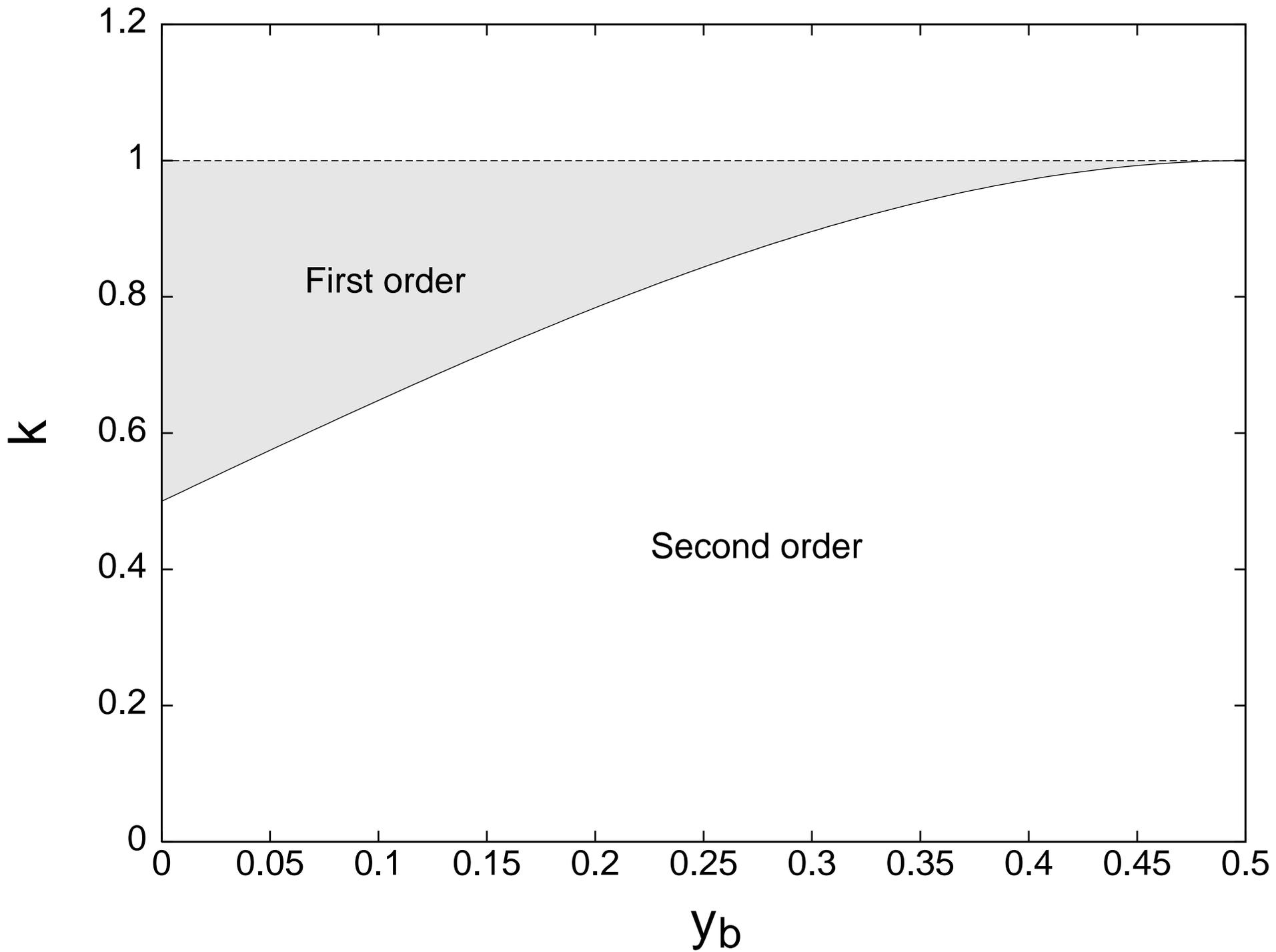
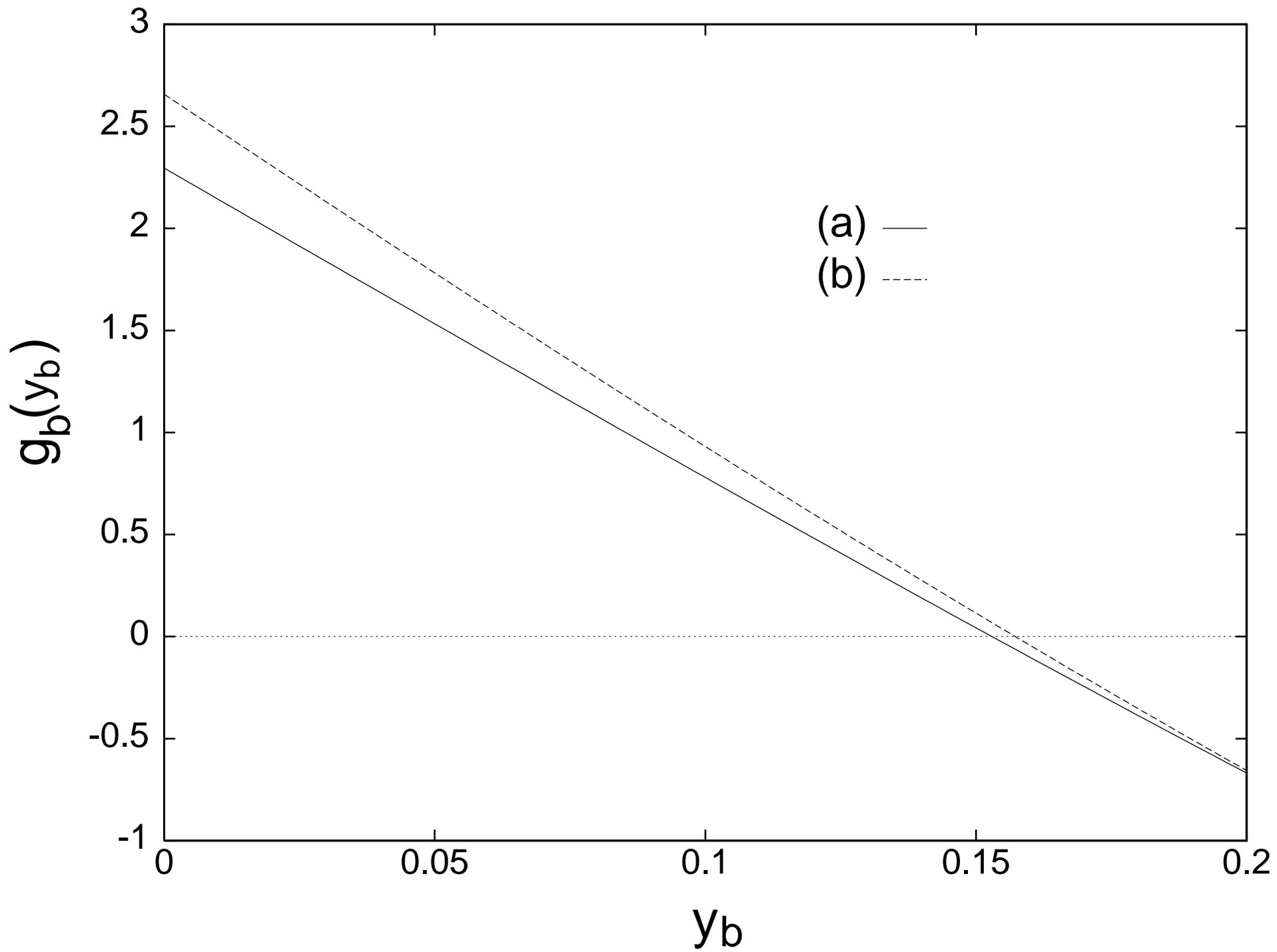
\end{document}